\documentstyle[fl eqn,twoside]{article}

\def\be{\begin{equation}}

\def\ee{\end{equation}}

\def\bi{\bibitem}

\begin{document}

\title{Scalar tensor theory of gravity carrying a conserved current}

\author{Abhik Kumar Sanyal}

\maketitle

\noindent

\begin{center}

Dept. of Physics, Jangipur College, Murshidabad,

\noindent

India - 742213\\

\noindent

and\\

\noindent

Relativity and Cosmology Research Centre\\

\noindent

Dept. of Physics, Jadavpur University\\

\noindent

Calcutta - 700032, India\\

\noindent

e-mail : aks@juphys.ernet.in\\

\end{center}

\noindent

\date{}

\maketitle
\begin{abstract}

A general scalar-tensor theory of gravity carries a conserved current for a trace free minimally coupled scalar field, under the condition that the potential $V(\phi)$ of the nonminimally coupled scalar field is proportional to the square of the parameter $f(\phi)$ that is coupled with the scalar curvature $R$. The conserved current relates the pair of arbitrary coupling parameters $f(\phi)$ and $\omega(\phi)$, where the latter is the Brans-Dicke coupling parameter. Thus fixing up the two arbitrary parameters by hand, it is possible to explore the symmetries and the form of conserved currents corresponding to standard and many different nonstandard models of gravity.  
\end{abstract}

\section{\bf{Introduction}}

The importance of scalar-tensor theory of gravity has always increased since the advent of the Brans-Dicke \cite{b:d} theory of gravity, which was originally introduced to incorporate Mach principle. Later, Brans-Dicke field has been found to arise even from higher dimensional theories, like superstring theories \cite{a:w}. Brans-Dicke theory leads to Einstein's theory in the limit $\omega \rightarrow \infty$ and so $\omega$, the Brans-Dicke function, is constrained by classical tests of general relativity. The light deflection and the time delay experiments demand $\omega > 500$, while the bounds on the anisotropy of the microwave background radiation demands $\omega < 30$. Hence, a viable model of the scalar-tensor theory of gravity requires $\omega$ to be a function of time and that too via the nonminimally coupled scalar field $\phi$. It has been observed that \cite{s:a} an asymptotic negative value of $\omega(\phi)$ leads to late time acceleration of the universe in Brans-Dicke cosmology. Induced theory of gravity \cite{a:m} on the other hand, appeared in an attempt to construct a gravitational theory consistent with quantum field theory in curved space time. It identifies the scalar field with the inflaton and has been found to be a strong candidate in several unified theories \cite{g:s}. Low energy effective action of string theory \cite{g:s} is also a nonminimally coupled scalar tensor theory of gravity, that contains a scalar field called dilaton. All these theories lead mostly to power law inflation \cite{k:k}. Further a scalar-tensor theory of gravity containing a coupling parameter in the form $f(\phi) = 1-\zeta\phi^2$, where, $f(\phi)$ is the parameter that is coupled with the scalar curvature $R$, is found to overcome the graceful exit problem and the problem of density perturbation for arbitrary large negative value of $\zeta$ \cite{f:u}. Recently, it has been observed \cite{a:d} that some string inspired scalar-tensor theories of gravitation \cite{g:v} lead to quintessence \cite{q:e}.  
\par
Due to such growing interests of scalar-tensor theories of gravitation, it is required to study the theory in some more detail.  All the above nonstandard theories of gravity which appeared in different context, contain the coupling parameters $f(\phi)$and $\omega(\phi)$ together with an arbitrary form of the potential $V(\phi)$. For a general form of scalar-tensor theories of gravity, these coupling parameters and the form of the potential are not known a-priori. It becomes less difficult to handle such theories if there exists some form of symmetry such that the corresponding conserved current somehow relates the coupling parameters. With this motivation, de Ritis et-al \cite{r:e} for the first time proposed that, if one demands the existence of Noether symmetry corresponding to such type of action, it might fix up the form of the coupling parameters and the potential. Indeed it does, if one of the two arbitrary parameters is considered. It has been observed \cite{a:b} however, that Noether symmetry often leads to some unpleasent features viz., it makes the Lagrangian degenerate, as the Hessian determinant $W = |\Sigma_{i,j}(\frac{\partial^2 L}{\partial\dot{q_{i}}\partial\dot{q_{j}}})| = 0$, and the effective Newtonian gravitational constant negative. Further, Noether theorem has So far been applied only in the cosmological models and it is not known how the existence of Noether symmetry for a nonstandard gravitational action can be claimed keeping the space-time arbitrary. Whatsoever, in the situations mentioned above \cite{a:b}, some other forms of symmetry were found to be present that are free from such unpleasent features. Those symmetries were found right from the field equations, rather from the invariance of the action under some suitable transformations, as is required to find Noether symmetry . 
\par
In view of the presence of such dynamical symmetries \cite{a:b} disussed above, here we are motivated to find such metric independent symmetry and the corresponding conserved current for a general form of scalar-tensor theory of gravity. In the following section we have achieved in finding such a conserved current for a trace free minimally coupled scalar field under the condition that the potential $V(\phi)$ of the nonminimally coupled field $\phi$ should be related to the parameter $f(\phi)$, that couples with the curvature scalar $R$. It is noteworthy that a trace free matter field corresponds to the radiation dominated era in the context of homogeneous cosmology. The conserved current found in the process relates the two coupling parameters considered here. Thus, such symmetry exists for all types of standard and nonstandard (scalar-tensor) theories of gravity. In section 3, we have made different choices of the coupling parameters and thus explored the forms of the conserved current in standard and some nonstandard models. Thus we have been able to study the form of symmetry of different theories of gravitation in a single frame-work. Section 4 is devoted to explore symmetry of axion-dilaton string effective action \cite{k:l}.            
\section{\bf{Action with a conserved current}}

The gravitational action nonminimally coupled to a scalar field $\phi$ can be expressed in the following general form,-

\be
A = \int[f(\phi) ~R -\frac{\omega(\phi)}{\phi}\phi,_{\mu}\phi^{,\mu}-V(\phi)-k(\frac{1}{2}\chi,_{\mu}\chi^{,\mu}+U(\chi))]\sqrt{-g}~d^4x,
\ee

where, $R$ is the Ricci scalar, $\sqrt{-g}$ is the determinant
of the metric of the four-space, $f(\phi)$ and $\omega(\phi)$ are coupling parameters while $\chi$ corresponds to a minimally coupled scalar field. The coupling constant $k=16\pi G$. The above action (1) corresponds to all types of scalar-tensor theories of gravity other than dilaton-axionic one \cite{k:l}, that we shall take up in section 4. The field equations corresponding to action (1) are,
\[
f(R_{\mu\nu}-\frac{1}{2}g_{\mu\nu}R)+f^{;\alpha}_{;\alpha}g_{\mu\nu}-f_{;\mu ;\nu}-\frac{\omega}{\phi}\phi_{,\mu}\phi_{,\nu}+
\frac{1}{2}g_{\mu\nu}(\frac{\omega}{\phi}\phi_{,\alpha}\phi^{,\alpha}+V(\phi))\]
\be
=\frac{k}{2}[\chi_{,\mu}\chi_{,\nu}-g_{\mu\nu}(\frac{1}{2}\chi_{,\alpha}\chi^{,\alpha}+U(\chi))]=\frac{k}{2}T_{\mu\nu}
\ee 
\be
Rf'+2\frac{\omega}{\phi}\phi^{;\mu}_{;\mu}+(\frac{\omega'}{\phi}-\frac{\omega}{\phi^2})\phi^{,\mu}\phi_{,\mu}-V'(\phi)
\ee
\be
\chi^{;\mu}_{;\mu}+\frac{dU}{d\chi}= 0
\ee
where,
\be
f^{;\mu}_{;\mu}=f''\phi^{,\mu}\phi_{,\mu}+f'\phi^{;\mu}_{;\mu}
\ee
In the above $T_{\mu\nu}$ is the energy-momentum tensor corresponding to the minimally coupled scalar field $\chi$ and dash $(')$ represents derivative with respect to $\phi$. Now the trace of equation (2) is, 
\be
Rf-3f^{;\mu}_{;\mu}-\frac{\omega}{\phi}\phi^{,\mu}\phi_{,\mu}-2V = \frac{k}{2}(\chi^{,\mu}\chi_{,\mu}+4U)=-\frac{k}{2}T^{\mu}_{\mu}
\ee
Hence we can eliminate the curvature scalar $R$, in view of equations (3) and (6) and thus obtain the following relation,
\be
(3f'^2+2f\frac{\omega}{\phi})'\phi^{,\mu}\phi_{,\mu}+2(3f'^2+2f\frac{\omega}{\phi})\phi^{;\mu}_{;\mu}+2f'V-fV'=-f'k(\chi^{,\mu}\chi_{,\mu}+4U)=f'k T^{\mu}_{\mu}
\ee
which can finally be expressed as 
\be
[(3f'^2+2f\frac{\omega}{\phi})^{\frac{1}{2}}\phi^{;\mu}]_{;\mu}-\frac{f^ 3}{2(3f'^2+2f\frac{\omega}{\phi})^{\frac{1}{2}}}(\frac{V}{f^2})'=-\frac{kf'}{2(3f'^2+2f\frac{\omega}{\phi})^{\frac{1}{2}}}(\chi^{,\mu}\chi_{,\mu}+4U)=\frac{kf'}{2(3f'^2+2f\frac{\omega}{\phi})^{\frac{1}{2}}}T^{\mu}_{\mu}
\ee
In view of the above equation (8) we can make the following statement that there exists a conserved current $J^{\mu}$, where
\be
J^{\mu}_{;\mu}=[(3f'^2+2f\frac{\omega}{\phi})^{\frac{1}{2}}\phi^{;\mu}]_{;\mu},
\ee
corresponding to the general form of scalar tensor theory of gravitational action (1), provided the trace of the energy-momentum tensor $T^{\mu}_{\mu}$ of the minimally coupled scalar field $(\chi)$ vanishes and the potential $V(\phi)$ for the nonminimally coupled scalar field $(\phi)$ is either zero or proportional to the square of the coupling parameter $f(\phi)$. Now, in the context of cosmology one breaks the general covariance and splits the space-time into space and time components by defining time-like hypersurfaces. Hence, one can transform the volume integral into the surface integral as,
\be
\int_{V} J^{\mu}_{;\mu}\sqrt{-g}d^4 x = \int_{\Sigma} J^{\mu}d\Sigma_{\mu}.
\ee
As a result, the above conservation principle implies that, on every time-like hypersurface $\int_{\Sigma} J^{\mu}d\Sigma_{\mu}=0$. Further, since 
\be
 J^{\mu}_{;\mu}=[(3f'^2+2f\frac{\omega}{\phi})^{\frac{1}{2}}\phi^{;\mu}]_{;\mu}=\frac{1}{\sqrt{-g}}[(3f'^2+2f\frac{\omega}{\phi})^{\frac{1}{2}}\phi^{,\mu}\sqrt{-g}]_{,\mu} = 0,
\ee
therefore, in the context of homogeneous cosmology,
\be
\frac{d}{dt}{J^{0}}= \frac{1}{\sqrt{-g}}\frac{d}{dt}[(3f'^2+2f\frac{\omega}{\phi})^{\frac{1}{2}}\dot \phi \sqrt{-g}] = 0,
\ee
where, dot represents time derivative and $J^{0}$ is the time component of the current density. Thus, the conservation principle demands that,
\be
(3f'^2+2f\frac{\omega}{\phi})^{\frac{1}{2}}\dot \phi \sqrt{g}
\ee 
is an integral of motion under the conition already stated, viz., $T^{\mu}_{\mu}= 0$ and $V(\phi)=\lambda f^2(\phi)$ or is zero, $\lambda$ being a constant. In the context of homogeneous cosmology, vanishing of the trace of the matter field either indicates radiation dominated era, ie. $\rho = 3p$ where, the energy density, $\rho = \frac{1}{2}\dot \chi^2 + U(\chi)$ and the pressure, $p = \dot \chi^2 - U(\chi)$, or the vacuum $\rho = p = 0$.

\section{\bf{Different forms of scalar tensor theory of gravity carrying a conserved current}}    
We have observed that the coupling parameters are not fixed in the process of finding the conserved current. There only exists a pair of relations amongst them, in view of the conserved current (9), which relates $\omega(\phi)$ with $f(\phi)$ and the relation between the $f(\phi)$ and the potential $V(\phi)$ in the form $V(\phi) = f^2(\phi)$. Therefore, fixing such parameters by hand it is possible to study different situations. This is done in this section. To understand the situations thus arise, we shall often refer to homogeneous cosmological models, viz., the isotropic Robertson-Walker space time given by

\be
ds^2=-dt^2+a(t)^2(d\chi^2+f^2(\chi)(d\theta^2+sin^2\theta~
d\phi^2)), 
\ee 
 and anisotropic axially symmetric Kantoski-Sachs space time given by
\be 
ds^2=-dt^2+a^2(t)dr^2+b^2(t)(d\theta
^2+sin^2 \theta ~d\phi ^2). 
\ee 

\subsection{Einstein's theory, $f =$ constant and $\omega = k\frac{\phi}{2}$.}
This case corresponds for $f = 1$ to the Einstein's theory of gravity with a pair of minimally coupled scalar fields $\phi$ and $\chi$, with the potential $V(\phi)= \lambda$, a constant. Thus the corresponding action  
\be
A = \int[R -k\frac{1}{2}\phi,_{\mu}\phi^{,\mu}-\lambda-k(\frac{1}{2}\chi,_{\mu}\chi^{,\mu}+U(\chi))]\sqrt{-g}~d^4x.
\ee  
carries a conserved current $J^{\mu}$, such that
\be
J^{\mu}_{;\mu}= \phi^{;\mu}_{;\mu} = 0,
\ee
in the absence of $\chi$-field or if the trace of the energy-momentum tensor corresponding to $\chi$-field vanishes ($T^{\mu}_{\mu} = 0$). In the context of homogeneous cosmology, the integral of motion is $\sqrt{g}\dot \phi$ which in the Robertson-Walker space-time reads $a^3 \dot \phi =$ constant. Since the potential $U(\chi)$ still remains arbitrary, so one has the liberty to consider the $\chi$-field as the quintessence field with a potential in the form $U(\chi) = \frac{\beta}{\chi^{\beta}}$, where, $\beta$ is a constant. Further, if the $\phi$-field is now treated as the perfect fluid source, then with the choice
\[\frac{1}{2}\dot \phi^2 + V = \rho_{\phi}\]
and
 \[ \frac{1}{2}\dot \phi^2 - V = p_{\phi} \]
where, $\rho_{\phi}$ and $p_{\phi}$ are the energy density and the pressure of the perfect fluid, we know that for $V = \lambda=0$ the above situation leads to stiff fluid equation of state $\rho_{\phi} = p_{\phi}$, for which we recover the well known result, viz., $\rho_{\phi} a^6=$ constant. On the other hand, if $V = \lambda \ne 0$, the same integral of motion $a^3 \dot\phi$ exists for the equation of state $\rho_{\phi} - p_{\phi} = 2\lambda$. Finally if $\dot \phi^2$ is small enough, ie., under slow roll approximation, $\lambda$ acts as cosmological constant for which the equation of state is $\rho_{\phi} + p_{\phi} = 0$.  

\subsection{Brans-Dicke theory, $f = \phi$ and $\omega =$ constant.}
In this situation $V = \lambda \phi^2$, and the action takes the form,
\be
A = \int[\phi R -\frac{\omega}{\phi}\phi,_{\mu}\phi^{,\mu}-\lambda \phi^2-k_{1}(\frac{1}{2}\chi,_{\mu}\chi^{,\mu}+U(\chi))]\sqrt{-g}~d^4x,
\ee  
which reduces exactly to the action for Brans-Dicke scalar tensor theory of gravity minimally coupled to a matter field, provided $V = \lambda \phi^2 = 0$. Equation (8) now takes the following form,
\be
 \phi^{;\mu}_{;\mu}= \frac{8\pi}{3+2\omega}T^{\mu}_{\mu}.
\ee
Thus we again recover the well known result and thus conclude that if the trace of the matter field vanishes, Brans-Dicke action admits a conserved current $J^{\mu}$, where, $J^{\mu}_{;\mu} = \phi^{;\mu}_{;\mu}= 0$, even in the presence of a potential in the form $V = \lambda \phi^2$ of the Brans-Dicke field. We remember that Brans-Dicke originally replaced $G$ by $\phi^{-1}$, hence, in the above action (18) we have deliberately introduced a new coupling constant $k_{1} = 16\pi$.

\subsection{Induced theory of gravity, $f = \epsilon \phi^2$ and $\omega = \frac{\phi}{2}$.}
Here, the potential is $V = \lambda \phi^4$ and the action is,
\be
A = \int[\epsilon \phi^2 R -\frac{1}{2}\phi,_{\mu}\phi^{,\mu}-\lambda \phi^4-k(\frac{1}{2}\chi,_{\mu}\chi^{,\mu}+U(\chi))]\sqrt{-g}~d^4x.
\ee  
The above action (20) for induced theory of gravity thus admits a conserved current $J^{\mu}$ such that, $J^{\mu}_{;\mu} = (\phi \phi^{;\mu})_{;\mu}= 0$, if the $\chi$-field is trace free, provided $\epsilon \ne -\frac{1}{12}$. In the context of homogeneous cosmology the conservation principle reads, 
\be
\sqrt{\epsilon(12\epsilon+1)}\frac{d}{dt}(\sqrt{-g}\phi \dot\phi) = 0.
\ee
In an attempt to find the forms of the coupling parameters and the potential by demanding the existance of Noether symmetry we have earlier observed that both in isotropic Robertson-Walker and anisotropic Kantowski-Sachs space-times \cite{a:b} Noether symmetry exists for $\epsilon = -\frac{1}{12}$, which makes the Newtonian gravitational constant negative and thus unphysical. It also makes the Lagrangian degenerate in the absense of the $\chi$-field, since Hessian determinant $W = |\Sigma_{i,j}(\frac{\partial^2 L}{\partial \dot q_{i}\partial \dot q_{j}})|$ vanishes. However, in that context, we have for the first time observed the existence of other conserved quantities found in view of the field equations, viz., $a^3 \phi \dot\phi =$ constant in the Robertson-Walker space-time and $ab^2 \phi \dot\phi =$ constant in anisotropic Kantowski-Sachs space-times, mentioned above. Thus, we recover the same results here too and hence conclude that induced theory of gravity admits above integral of motions given in (21), in vacuum or in the radiation dominated era provided, the induced field has got a potential in the form, $V(\phi) = \lambda \phi^4$.

\subsection{String effective action $(3f'^2+2f\frac{\omega}{\phi})' = \beta(3f'^2+2f\frac{\omega}{\phi})$.}
Under the above assumption, $f(\phi)$ and $\omega(\phi)$ are related in the following manner,
\be
 3f'^2+2f\frac{\omega}{\phi}=\alpha e^{\beta \phi},
\ee
where, $\alpha$ and $\beta$ are constants. Thus, there exists a conserved current $J^{\mu}$, such that
\be
J^{\mu}_{;\mu} = (e^{\frac{\beta}{2}\phi}\phi^{;\mu})_{;\mu} = 0.
\ee
One can fix up $\omega(\phi)$ by choosing $f(\phi)$ in the form
\[
f(\phi) = n e^{\frac{\beta}{2}\phi}
\]
where, $n$ is a constant. As a result, $\omega(\phi)$ takes the form
\[
\omega(\phi) = \frac{2\alpha-3n^2 \beta^2}{4n}\phi e^{\frac{\beta}{2}\phi}.
\]
Thus, the corresponding action is,
\be
A = n\int[e^{\frac{\beta}{2}\phi}\{ R -\frac{2\alpha-3\beta^2 n^2}{4n^2}\phi,_{\mu}\phi^{,\mu}-\Lambda e^{\frac{\beta}{2} \phi}\}-k(\frac{1}{2}\chi,_{\mu}\chi^{,\mu}+U(\chi))]\sqrt{-g}~d^4x.
\ee  
For $\beta > 0$ and a decaying $\phi$-field, the above action in the cosmological context, asymptotically goes over to the Einstein's action with a cosmological constant, which is minimally coupled to a scalar field $\chi$, that acts as a perfect fluid source. Thus the action (24) asymptotically leads to de-Sitter universe in theabsence of the $\chi$-field. However, the action (24) reduces to the Gravi-dilaton string effective action minimally coupled to the scalar field $\chi$, for $\beta = -2$ and $\alpha = 4n^2$, given by Gasperini and Veneziano \cite{a:d} as, 
\be
A = n\int[e^{-\phi}\{ R -\phi,_{\mu}\phi^{,\mu}-\Lambda e^{- \phi}\}-k(\frac{1}{2}\chi,_{\mu}\chi^{,\mu}+U(\chi))]\sqrt{-g}~d^4x.
\ee  

In the above action (25), $n = \frac{M^2_{s}}{2}$ is the fundamental string length parameter. Hence, the Gravi-dilaton string effective action admits an integral of motion $a^3 e^{-\phi}\dot \phi^2$ in the Robertson-Walker space-time and $ab^2 e^{-\phi}\dot \phi^2$ in Kantowski-Sachs space-time in the radiation dominated era. It is to be mentioned that Gasperini \cite{a:d} has shown that such an action leads to quintessential effects.
\par
Late time acceleration of the universe may also be realized in view of the following action, different from the above Gravi-dilaton string effective action, carrying the same conserved current (23), viz.,
\be
A = \int[\phi^n R -(\frac{\alpha e^{\beta\phi}}{2\phi^n}-\frac{3n^2}{2}\phi^{(n-2)})\phi,_{\mu}\phi^{,\mu}-\Lambda \phi^{2n}-k(\frac{1}{2}\chi,_{\mu}\chi^{,\mu}+U(\chi))]\sqrt{-g}~d^4x,
\ee
where, we have chosen $f(\phi) = \phi^n$. Thus, $V(\phi) = \Lambda \phi^{2n}$, and
\be
\omega=\frac{\alpha e^{\beta\phi}-3n^2 \phi^{2(n-1)}}{2\phi^{(n-1)}}.
\ee
The above form (27) of $\omega(\phi)$ admits a sign flip and so $\phi$-field acts as exotic matter. For $n = 0$, $\omega = \frac{\alpha}{2}\phi e^{\beta\phi}$, and hence there is no sign flip. However, for $n = 1$, $\omega = \frac{\alpha}{2}
e^{\beta\phi} - \frac{3}{2}$ and for $n = 2$, $\omega = \frac{\alpha}{2}
\frac{e^{\beta\phi}}{\phi} - 6\phi$, and so on. For large positive $\beta$ and for a decaying $\phi$-field, the first term of the last two expressions of $\omega$ falls of rapidly and hence $\omega < 0$ asymptotically, as a result, late time acceleration may be realized. For $n > 1$, $\omega$ finally vanishes, which may cause a future deceleration of the universe. The result is more pronounced for $n = -1$, for which $\omega = \frac{\alpha}{2}\phi^2 e^{\beta\phi} - \frac{3}{2\phi^2}.$    
\subsection{Nonminimal coupling $3f'^2+2f\frac{\omega}{\phi} = f^2_{0} =$ constant.}
Despite the standard forms of the scalar tensor theory of gravity investigated so far, we can also study many other nonminimally coupled theories carrying a conserved current. Some of these are explored in this subsection. 
\par
Under the above choice the conserved current $J^{\mu}$ in view of equation (9) is such that 
\be
J^{\mu}_{;\mu} = \phi^{;\mu}_{;\mu} = 0.
\ee
We shall now choose either $\omega(\phi)$ or $f(\phi)$ to fix up the other along with the potential $V(\phi)$.
\par
{\bf{Case 1. $\omega = \frac{\phi}{2}$}.}
\par
As a result of such a choice,
\be
f(\phi) = f_{0}^2 - \frac{\phi^2}{12}, ~~~ V(\phi) = (f_{0}^2 - \frac{\phi^2}{12})^2.
\ee
Thus, in the cosmological context, if $\phi$ falls off with time then the signature of the effective gravitational constant flips from a negative to positive value. To restrict $f$ to a positive value throughout, $\phi^2$ should be restricted to, $\phi^2 < 12 f_{o}^2$.
\par
{\bf{Case 2. $f = \phi^n, n$}} being a constant.
\par
Hence, 
\be
V(\phi) = \lambda \phi^{2n},~~~\omega = \frac{f_{0}^2-3n^2 \phi^{2(n-1)}}{2\phi^{(n-1)}}
\ee
Now, for $n = 0$, $f = 1, V = \lambda =$ constant, and $\omega = \frac{f_{0}^2}{2}\phi$. Thus for $f_{0}^2 = k$, we recover Einstein's gravity with a pair of nonminimally coupled fields, which has been studied in subsection 3.1.
\par
For, $n = 1$, $f(\phi) = \phi, V(\phi) = \lambda\phi^2$ and $\omega = \frac{f_{0}^2 - 3}{2}$ and hence we recover Brans-Dicke theory of gravity, which has been studied in subsection 3.2.
\par
For $n = 2$, $f(\phi) = \phi^2, V(\phi) = \lambda\phi^4$ and $\omega = \frac{f_{0}^2 - 12\phi^4}{2\phi}$. In this situation, a sign flip of $\omega(\phi)$ from negative to positive value for decaying $\phi$ field is observed and as a result, in the cosmological context, $\omega(\phi)$ turns out to be indefinitely large asymptotically. For, higher values of $n$ situation does not alter and the qualitative features remain the same.
\par
For $n = -1$, $f(\phi) = \phi^{-1}, V(\phi) = \lambda\phi^{-2}$ and $\omega = \frac{f_{0}^2\phi^4 - 3}{2\phi^2}$.
\par
Thus, the action (1) now takes the following form, 
\be
A = \int[\frac{R}{\phi}-\omega\phi_{,\mu}\phi^{,\mu}-\lambda\phi^{-2}-L_{m}]\sqrt{-g}d^4x,
\ee
where, $L_{m}$ indicates matter lagrangian. This action has got some interesting features. In the context of homogeneous cosmology, this action admits an integral of motion $\sqrt{-g}\dot\phi$, in the radiation era. The form of the potential $V(\phi)$ indicates that the $\phi$-field acts as 'Quintessent' field. Indeed it is so, since for a time decaying $\phi$-field the sign of $\omega$ flips from a positive to negative value. Further, $\frac{R}{\phi}$ tends to remain finite asymptotically. Thus it appears that late time acceleration of the Universe can be explained in view of the above action.
\section{\bf{Axion-Dilaton string effective action carrying a conserved current.}}
To study the symmetry of axion-dilaton string effective action we consider the following general form of such an action, 
\be
A = \int[f(\phi) ~R -\frac{\omega(\phi)}{\phi}\phi,_{\mu}\phi^{,\mu}-V(\phi)-h(\phi)\chi,_{\mu}\chi^{,\mu}+U(\chi)]\sqrt{-g}~d^4x,
\ee
where, we have introduced yet another coupling parameter $h(\phi)$. The above action reduces to the axion-dilaton string action for $f = $ constant and $\frac{\omega}{\phi} = \frac{1}{2}$.  Proceeding in a similar fashion as is done in section 2, we can construct the following equation that is complementary to equation (8), viz.,
\be
[(3f'^2+2f\frac{\omega}{\phi})^{\frac{1}{2}}\phi^{;\mu}]_{;\mu}-\frac{f^ 3}{2(3f'^2+2f\frac{\omega}{\phi})^{\frac{1}{2}}}(\frac{V}{f^2})'=-\frac{[\chi^{,\mu}\chi_{,\mu}(hf'-h'f)+4Uf']}{2(3f'^2+2f\frac{\omega}{\phi})^{\frac{1}{2}}}.
\ee
 We thus make the following statement that there exists a conserved current $J^{\mu}$ corresponding to the action (32), such that,
\be
J^{\mu}_{;\mu}=[(3f'^2+2f\frac{\omega}{\phi})^{\frac{1}{2}}\phi^{;\mu}]_{;\mu},
\ee
provided, 
\[ V(\phi) = \lambda f^2 (\phi),~~h(\phi) = k f(\phi),~~U(\chi) = 0, \]  
where, $\lambda$ and $k$ are constants. The action that admits such a conserved current, is given by,
\be
A = \int[f(\phi) ~R -\frac{\omega(\phi)}{\phi}\phi,_{\mu}\phi^{,\mu}-\lambda f^2(\phi)-k f(\phi)\chi,_{\mu}\chi^{,\mu}]\sqrt{-g}~d^4x,
\ee
Hence, we observe that a conserved current exists here too but under much restrictive condition. It is noteworthy that all the coupling parameters $h(\phi), \omega(\phi)$, and the potentials $V(\phi), U(\chi)$ are now fixed, once $f(\phi)$ is fixed up by hand. However, it is not possible to recover dialton-axionic form of the action, since, for $f =$ constant, $h(\phi)$ also becomes a constant, and $\chi$-field turns out to a minimally coupled one. Therefore we understand that dilaton-axion string field action does not admit a symmetry in the classical context, though it admits some symmetry \cite{j:m} in the quantum level. Nevertheless, it is interesting to study this situation also. In the context of homogeneous cosmology there exist an integral of motion $g \dot \phi^2 \frac{\omega}{\phi}$ under this situation, and thus $\phi$-field still behaves as a nonminimally coupled one.

\section{\bf{Concluding remarks}}  
The existence of a general form of symmetry and the corresponding conserved current for the standard and some nonstandard theories of gravity has been found for a trace free minimally coupled scalar field under the condition that the potential and one of the coupling parameters are related as $V(\phi) = f^2 (\phi)$. The conserved current relates the two coupling parameters $f(\phi),~ \omega(\phi)$ with the scalar field $\phi$ and the scale factor, in the context of homogeneous cosmology. Thus choosing the coupling parameters by hand we have explored symmetries of different standard and nonstandard models. The existence of the integral of motion in the cosmological context makes it easier to handle the field equation for studying exact solutions. The symmetry thus found is not a result of the invariance of action under some suitable transformation and so is not an artifact of Noether's theorem. It has been also observed that dilaton-axionic string field action does not admit such symmetry.

\end{document}